# A deep learning framework for the detection and quantification of drusen and reticular pseudodrusen on optical coherence tomography

Roy Schwartz[1,2,3], Hagar Khalid[1], Sandra Liakopoulos[4,5], Yanling Ouyang[1], Coen de Vente[3,6,7], Cristina González-Gonzalo[3,7], Aaron Y. Lee[8], Robyn Guymer[9], Emily Y. Chew[10], Catherine Egan[1], Zhichao Wu[9], Himeesh Kumar[9], Joseph Farrington[2], Clara I. Sánchez[3,6] and Adnan Tufail[1]

[1] Moorfields Eye Hospital NHS Foundation Trust, London, UK
[2] Institute of Health Informatics, University College London, London, UK
[3] Quantitative Healthcare Analysis (qurAI) Group, Informatics Institute, University of Amsterdam, Amsterdam, Netherlands
[4] Cologne Image Reading Center, Department of Ophthalmology, Faculty of Medicine and University Hospital Cologne, University of Cologne, Cologne, Germany
[5] Department of Ophthalmology, Goethe University, Frankfurt, Germany
[6] Amsterdam UMC location University of Amsterdam, Biomedical Engineering and Physics, Amsterdam, Netherlands
[7] Diagnostic Image Analysis Group (DIAG), Department of Radiology and Nuclear Medicine, Radboudumc, Nijmegen, Netherlands
[8] Department of Ophthalmology, University of Washington, Seattle WA
[9] Centre for Eye Research Australia, Royal Victorian Eye and Ear Hospital, East Melbourne, Australia; Ophthalmology, Department of Surgery, The University of Melbourne, Melbourne, Australia.
[10] National Eye Institute (NEI), National Institutes of Health (NIH), Bethesda, MD USA

Corresponding author:
Roy Schwartz
162 City Road,
EC1V 2PD, London, UK
royschwartz@gmail.com

Abstract

**Purpose**: To develop and validate a deep learning (DL) framework for the detection and quantification of drusen and reticular pseudodrusen (RPD) on optical coherence tomography scans.

**Design**: Development and validation of deep learning models for classification and feature segmentation.

**Methods**: A DL framework was developed consisting of a classification model and an out-of-distribution (OOD) detection model for the identification of ungradable scans; a classification model to identify scans with drusen or RPD; and an image segmentation model to independently segment lesions as RPD or drusen. Data were obtained from 1,284 participants in the UK Biobank (UKBB) with a self-reported diagnosis of age-related macular degeneration (AMD) and 250 UKBB controls. Drusen and RPD were manually delineated by five retina specialists. The main outcome measures were sensitivity, specificity, area under the ROC curve (AUC), kappa, accuracy and intraclass correlation coefficient (ICC).

**Results**: The classification models performed strongly at their respective tasks (0.95, 0.93, and 0.99 AUC, respectively, for the ungradable scans classifier, the OOD model, and the drusen and RPD classification model). The mean ICC for drusen and RPD area vs. graders was 0.74 and 0.61, respectively, compared with 0.69 and 0.68 for intergrader agreement. FROC curves showed that the model's sensitivity was close to human performance.

**Conclusions**: The models achieved high classification and segmentation performance, similar to human performance. Application of this robust framework will further our understanding of RPD as a separate entity from drusen in both research and clinical settings.

Introduction

Age-related macular degeneration (AMD) is defined by the presence of drusen, deposits found under the retinal pigment epithelium (RPE), which are key to the diagnosis of AMD.[1] Recent advances in multimodal imaging have, however, allowed us to substantially improve our ability to characterize the AMD phenotype, revealing information about a variety of the deposits that occur in AMD, such as reticular pseudodrusen (RPD).[2] RPD have been associated with late AMD and are considered a critical AMD phenotype to understand.[3–12] To date, most studies associating AMD risk with RPD have relied on a binary presence of RPD (i.e., their presence or absence) with no clear understanding of how the quantity of RPD plays into the risks posed by their presence. Understanding associations and risk of RPD is confounded by the fact that eyes with RPD often also have drusen which impose their own risks. To help improve our understanding of RPD and their associations, large datasets are essential but to date most available large datasets are based on cohorts collected for their AMD status, and few have eyes with only RPD. This leads to confounder issues when trying to understand the contribution that RPD make to any increased risk of vision loss in eyes with AMD.

Spectral-domain optical coherence tomography (SD-OCT) has been shown to have a much higher sensitivity and specificity for both detecting RPD and separating lesions from drusen compared with the blue channel of color fundus photographs (CFP), infrared reflectance, fundus autofluorescence, near-infrared fundus autofluorescence, confocal blue reflectance, and indocyanine green angiography.[13,14] In addition, OCT is the only imaging modality that allows the confirmation of the subretinal localization of RPD, which cannot be ascertained by other imaging modalities.[2] Given the subtlety of RPD lesions on OCT, even on latest generation devices, and more so on early generation OCT utilized in existing large population studies, human detection and quantification remain a challenge.[15] Given the importance of being able to detect and quantify RPD and separate them from drusen in terms of both our understanding of the pathogenesis of RPD and the potential implication of their presence in current and future therapies,[16] an automated approach to classification and quantification is needed.

Machine learning (ML) algorithms have been shown to be powerful tools in the automatic quantification of retinal biomarkers identified on OCT,[17–19] making them ideal for the detection of RPD and drusen. To date, there is a large volume of published studies describing the detection of drusen on OCT using ML, the majority of which deploy classification models which do not allow for the quantification of lesion area.[20–24] Thus far, only two studies explored ML techniques for the automatic detection of RPD on OCT. The first was a classifier, thus not allowing for image quantification,[25] and the other was based on the identification of drusen and RPD by interpolating retina layer undulations. The latter approach was only internally validated on a small number of eyes and has not been shown to perform on images from the more challenging imaging generated from older SD-OCT devices used in a number of large population studies or distinguish between RPD stages.[26]

We herein present a deep learning (DL) framework for the detection and quantification of drusen and RPD in the UK Biobank (UKBB), a large-scale biomedical database and research resource containing genetic, lifestyle and health information from half a million UK participants.

Methods

**Study population**

The UKBB study is a large, multisite, community-based cohort study with the aim of improving the prevention, detection, and treatment of a wide range of serious and life-threatening diseases. UKBB's database includes data on 500,000 volunteer participants aged between 40-69 years, recruited in 2006-2010 from across the UK. All UK residents aged 40 to 69 years who were registered with the National Health Service and living up to 25 miles from one of 22 study assessment centers were invited to participate. The North West Multi-centre Research Ethics Committee approved the study (REC reference number: 06/MRE08/65), in accordance with the principles of the Declaration of Helsinki. Detailed information about the study is available at the UKBB website ([www.ukbiobank.ac.uk](www.ukbiobank.ac.uk)).

Of all participants in the UKBB, 67,687 participants underwent OCT and CFP imaging, at six UKBB centers (Sheffield, Liverpool, Hounslow, Croydon, Birmingham, and Swansea) acquired using the Topcon 3D OCT 1000 Mark II (Topcon, Japan). Image acquisition was performed under mesopic conditions, without pupillary dilation, using the 3-dimensional macular volume scan (512 horizontal A-scans/B-scan; 128 B-scans in a 6x6-mm raster pattern). Of 2,622 participants with a self-reported diagnosis of AMD identified in the database, 1,284 had OCT volume scans and CFP and were used in the study. The UKBB project ID associated with this paper is 60078. Patients were excluded from the study if they had withdrawn their consent.

**Deep Learning Framework**

Upon visual inspection, a significant number of OCT scans were found to be of insufficient quality for this study. To mitigate this, and to improve the accuracy of DL, a framework consisting of several separate DL models was developed (Figure 1): a. A classification model to detect ungradable scans (Ungradable Classification Model), based on the difference in signal-to-noise ratio between gradable and ungradable scans. b. An out-of-distribution detection model to further classify ungradable scans (see Model Development below), based on the difference between gradable and ungradable scans resulting from outliers caused by optical artifacts (Outlier Detection Model). c. A classification model to identify scans with drusen or RPD vs. controls (those without these lesions) (Drusen/RPD Classification Model). d. An image segmentation model to independently segment lesions as RPD or drusen, allowing their quantification (Drusen/RPD Segmentation Model).

**Data selection**

*Classification models*

To train the classification models, each OCT volume (eye) was labeled by a single grader (R.S.) as ungradable; containing drusen/RPD, or both; or control (not containing drusen or RPD). Volumes were deemed ungradable if the outer retina was not clearly seen in a scan in a manner that would allow to confirm or reject the presence of RPD and drusen (e.g., due to image noise, shadowing, or clipping of the outer retina) or in cases where vertically flipped scans existed in the volume.

Drusen were defined as discrete areas of RPE elevation with low to medium reflectivity, similar to the reflectivity of the inner plexiform and ganglion cell layers. RPD were defined as lesions above the RPE with medium reflectivity, similar or slightly less than the reflectivity of the retinal nerve fiber layer. Each of the previously described stages were also considered when labeling eyes as RPD: [2,27] Stage 1 - diffuse deposition of granular hyperreflective material between the RPE and the ellipsoid zone (EZ); Stage 2 - similar to stage 1, but the mounds of accumulated material are sufficient to alter the contour of the EZ, resulting in EZ undulations; Stage 3 - the material is thicker, adopts a conical appearance, and breaks through the EZ; Stage 4 - defined by fading of the material because of reabsorption and, eventually, migration within the inner retinal layers.

Each eye was graded according to the following scale: 1. No drusen/RPD 2. One drusen/RPD; 3. More than one drusen/RPD; 4. Questionable drusen/RPD; 5. Ungradable. Categories 1 and 3 were used to train the drusen detection classification model, and category 5 was used to train the ungradable detection classification model. Category 2 was not used since the identification of RPD is challenging, and the presence of a pattern helps to distinguish cases with genuine RPD versus human variability. Therefore, to reduce the risk of including false-positive cases, it was decided to include only cases with more than a single RPD lesion. For uniformity, the same was applied to drusen. Category 4 was not used as the inclusion of questionable lesions might degrade the model's performance.

Of 2,622 participants with self-reported AMD, 1,284 had OCT scans. Four hundred and eighty-nine eyes of 287 patients were classified as having more than one drusen; 57 eyes of 38 patients were classified as having more than one RPD; 343 eyes of 232 patients were classified as ungradable; 1,182 eyes of 591 patients were identified as having no drusen/RPD (controls). In addition, to avoid selection bias that may result from the selection of controls out of a population of self-reported AMD, 250 control eyes were randomly identified from the general cohort. Eventually, 500 control eyes, 468 eyes with any drusen, or RPD and 308 eyes with ungradable scans were included. They were divided into a training, validation, and test set by a ratio of 60:20:20. Eyes of a specific participant were not allowed to exist in more than one set.

*Semantic segmentation model*

To train the semantic segmentation model, additional cases were identified in the UKBB outside of the cohort of participants with a self-reported diagnosis of AMD using an in-house available deep learning approach developed to detect AMD features in CFPs. It does so in a hierarchical manner by first detecting drusen. Of those, it then detects large drusen, and of those it detects

RPD.[28] By using this approach and manually removing low-quality images, an additional 22 eyes were found to have more than one RPD after visual inspection and included in the training set for the segmentation model

As the model was trained on B-scans rather than OCT volumes, B-scans were classified by a single grader (R.S.) into the three groups as previously mentioned: 2,834 scans with RPD, 2,338 with drusen, and 4,946 controls. Of those, B-scans with RPD were selected manually for training if they contained at least one RPD, with or without drusen, from different areas of the macula, to reflect the variability in RPD appearance. Overall 334 B-scans with RPD were included. The same number of B-scans of drusen and controls were randomly selected for training. These were divided into a training, validation, and test set (using a ratio of 60:20:20). B-scans of a specific participant were not allowed to exist in more than one set.

**Annotation**

Manual delineation of features (drusen and each stage of RPD) to train the image segmentation model was performed by five experienced graders. The training and validation sets were independently annotated by two retina specialists (R.S., H.K.) and the second grader (H.K.) was used as the ground truth for training the model. An additional three retina specialists (A.T., S.L., Y.O) independently annotated all the scans in the test set. Annotation was done using Label Studio version 1.2.[29] The graders were provided a list of B-scans, shuffled to avoid priming bias (i.e., the tendency to annotate lesions based on previously seen lesions in the same eye). They had access to the complete OCT volume and could zoom in for accurate delineation. A document containing instructions and examples of the correct annotation of labels of interest was provided to graders and discussed with them. It included the definitions mentioned previously for drusen and different stages of RPD. Each of the lesion types was assigned a label and a different color. Graders were asked to grade a standard set of 6 B-scans containing examples of each label prior to annotating their respective sets and an adjudication process took place (R.S.) to ascertain uniformity among graders.

**Model development**

All models were trained on a single server with an Intel 18 core 4.6 GHz Xeon processor, 256 GB of RAM, and an Nvidia Quadro RTX8000 card with 48 GB of RAM.

*Classification models*

The architecture for the Ungradable Classification Model and the Drusen/RPD Classification Model was a 3D Inception-V1.[30] 2D convolutions in the original Inception-V1 model were replaced with 3D convolutions. Except for the last convolution, a batch normalization layer [31] and rectified linear unit (ReLU) activation function [32] followed each convolution. The last convolution was followed by a softmax layer. We used Adam [33] with a learning rate of 10-4, $\beta_1 = 0.9$ and $\beta_2 = 0.999$ as the optimizer. During training, batches were randomly sampled in a balanced manner such that samples from each class were chosen equally often. Cross-entropy was used for the loss function. We employed early stopping with a patience of 10,000 iterations based on the kappa score on the validation set. Data augmentation was applied to the training set, which consisted of random rotations between -20 and +20 degrees, shearing between -10% and +10%, zooming between -10% and +10%, translations between -10 and +10 pixels in the B-scan plane, translation between +2 and -2 pixels in the z-direction, horizontal B-scan flipping with a probability of 15%, gaussian noise with a mean of 0, a standard deviation of 0.1 and a probability of 15%, gamma corrections with γ between 0.75 and 3.0 and a probability of 15%.

The Ungradable Classification Model was trained on a specific dataset, as mentioned above, which involved specific types of image aberrations. Since the data used to train the model represents only roughly 1.5% of the total UKBB dataset, a model trained to identify specific types of aberrations might not generalize well to the whole dataset (or other datasets). Therefore, as part of the ungradable detection algorithm, we used deep ensembles [34] for out-of-distribution (OOD) detection in addition to the previously mentioned classification model. This is a commonly used technique for uncertainty estimation and OOD detection that approximates Bayesian neural networks. Thus, it should detect any deviation from normal scans, which should in theory also identify aberrations the previous model was not trained on. In this work, the deep ensemble consisted of 10 individual models, each individually trained on the entire training set with different weight initializations and different seeds for random sampling. During inference, we used the mean variance for each class among the models in the ensemble as a measure for the uncertainty of a sample. Ungradable cases were then differentiated from gradable ones based on this uncertainty measure. Of note, both models were tested on the totality of the test set.

*Semantic segmentation model*

A 2D U-Net architecture [35] was trained using the nnU-Net framework, which has achieved high-performance values for various medical segmentation tasks and has the advantage of automatically adapting to different biomedical datasets.[36] For training, 5-fold cross-validation was used and testing was performed with an ensemble of the cross-validated models.

Due to the limited numbers of B-scans for stage 3 and 4 RPD, stages 2, 3, and 4 RPD were grouped together as a single class. Thus, the model was trained to distinguish between three classes: drusen; stage 1 RPD; and stages 2, 3, and 4 RPD.

**Statistical analysis**

We evaluated the performance of the classification models using five metrics defined as follows: a. Area under the Receiver Operating Characteristic (ROC) curve (AUC) - an ROC curve [37] displays the trade-off between the true-positive rate and true-negative rate of a classification model at different threshold levels. AUC represents the model's capability to separate the negative and positive classes; b. Accuracy - the percentage of correctly classified images. c. Cohen's kappa [38] - compares the observed accuracy with an expected accuracy (random chance); d. Sensitivity; e. Specificity; f. Area under the precision-recall curve – a precision-recall curve displays the trade-off between the positive predictive value and the sensitivity of a classification model at different threshold levels.

We evaluated the performance of the segmentation model using the following measures: To measure the segmentation performance we identified the number of individual features that were properly detected (i.e., overlapped with the ground-truth segmentation of the feature) within each B-scan by using the Label function of the Scikit Image Python library, which finds connected components in a binary image. We analyzed the overlap using free-response receiver operating characteristic (FROC) curves. We also reported the Dice similarity metric, which is defined as the size of the intersection of 2 areas divided by their average individual size. A Dice-score of 1 indicates perfect agreement and a score of 0 indicates disjoint areas. [39]

In addition, the intraclass correlation coefficient (ICC) for absolute agreement was used to measure agreement in the area of the different lesions between the model and graders and for interrater reliability analysis. For model-grader agreement, the mean of the ICC between the model's segmented areas and those segmented by each grader is presented. Cases with no segmentation were included as zero area. The ICC was calculated using the Pingouin library for Python.[40] ICC values were interpreted as follows: [41] A value below 0.50 was considered poor; a value between 0.50 and 0.75 was considered moderate; a value between 0.75 and 0.90 was considered good; and a value above 0.90 was considered excellent.

Results

*Classification models*

Metrics obtained by the different models are presented in Table 1, and receiver operating characteristic (ROC) curves are presented in figure 2. Each model was tested on the totality of the test set. All models achieved a high AUC, ranging from 0.93 to 0.99, and high accuracy, ranging from 81.6% to 98.4%. Between the models aimed at image quality assessment, the Ungradable Classification Model achieved a higher sensitivity, while the Outlier Detection Model achieved high sensitivity. The Kappa scores ranged from 0.59 to 0.97. Precision-recall curves for the different models are presented in Supplementary Figure 1.

*Segmentation model*

Quantitative results for the ICC for each feature are presented in Table 2. The ICC for the model's performance against all graders was averaged and is presented alongside the model-grader performances. In addition, the intergrader agreement between all three graders is presented. When considering the test set, for drusen, the model and graders both achieved moderate agreement, with higher agreement achieved by the model compared to the intergrader agreement. For stage 1 RPD, both the model and human graders achieved poor agreement, again with the model exceeding human agreement. The agreement of both humans and model was again poor for stages 2, 3, and 4 RPD, this time with intergrader agreement exceeding model agreement. When the RPD area of all RPD stages combined was considered, both the intergrader agreement and the model's agreement with the ground truth were moderate, with the intergrader agreement exceeding the model'.

FROC curves comparing the model's performance against each grader are presented in Figure 3. The most experienced grader for this task was chosen as a reference standard against model performance and against other graders. This figure highlights the sensitivity for both the graders and the model when operating at varying false-positive rates, with confidence intervals obtained by bootstrapping (1000 bootstrap samples). For drusen, stage 1 RPD, stage 2, 3, and 4 RPD, and all stages RPD, the 95% confidence interval of the model overlaps with the confidence interval of the grader marked in blue, and only for drusen with both graders. For drusen, for stages 2, 3, and 4 RPD, and for all stages RPD the model obtained a sensitivity that is lower than both graders when operating at the same false-positive rate, while for RPD stage 1 it was higher than one grader and lower than the other.

The Dice scores between the model and graders and between grader pairs is presented in Supplementary Table 1. Qualitative results of the output of the segmentation model are shown in Figures 4, 5, and 6.

Discussion

We present a robust deep learning framework for the elimination of ungradable scans, classification of drusen, and segmentation of drusen and RPD. To the best of our knowledge, this is the first framework handling different aspects of lesion analysis in AMD, including automated image quality assessment and lesion detection, and this is the first DL model to allow accurate quantification of these lesions.

Our two classifiers for image quality assessment were designed to perform two different tasks: the first, quality assessment, achieved by detecting the difference in signal-to-noise ratio between gradable and ungradable scans; and the second - detection of outliers caused by optical artifacts by OOD detection. Both classifiers achieved high performance in detecting poor quality scans (AUC of 0.95 and 0.93 for the Ungradable Classification Model and the Outlier Detection Model, respectively). They both serve as steps in automated data curation. Image quality control is essential to ensure optimal performance by a DL algorithm designed to be deployed on real-world data.[42] Unlike research and development environments, where such models are often trained on carefully curated datasets, real-world data may be more challenging, as evidenced by a recent attempt by Google Health to deploy a diabetic retinopathy model in a clinic setting, where its performance was worse than in the lab setting.[43] To date, only a small number of publications described the use of ML for image quality assessment on OCT scans. For example, Kauer et al. developed an ML classifier (AQuA), which was trained on OCT images acquired on the Spectralis SD-OCT device (Heidelberg Engineering) to identify poor quality scans.[44] Later, another neural network termed AQuANet was developed to allow AQuA to be adapted to OCT devices from other vendors. It was shown to transfer well to the Cirrus HD-OCT device (Carl Zeiss Meditec AG).[45] However, both devices are characterized by high-quality scans which are often lacking in existing large population studies' datasets acquired on older devices. To the best of our knowledge, our framework is the first published to classify poor quality scans on such devices, making it useful for research involving similar large datasets. It is also the first to utilize image quality control as part of a detection and quantification framework, a fact which should increase its accuracy when deployed on target datasets. The use of out-of-distribution detection alongside a classifier trained on specific examples of ungradable volumes allows the model to be more generalizable to previously unseen image artifacts.

The Drusen/RPD Classification Model achieved an AUC of 0.99. To the best of our knowledge, this is the first classifier that can detect both drusen and RPD. Numerous studies previously reported on classification algorithms for the detection of drusen only.[20–22,46] The ability to detect RPD as well as drusen can be used both for screening high-risk patients and for research into the latter, in addition to its role in the current framework.

For the Drusen/RPD Segmentation model, the model's agreement with human graders, as reflected by ICC scores, was better than the intergrader agreement for drusen. In regards to RPD, it was better for stage 1 RPD. Stage 1 RPD, as reflected by the very low intergrader agreement for segmentation of this lesion, is an exceptionally challenging lesion to grade since it only presents as a medium reflectivity change between the RPD and EZ with the additional loss of the normal anatomy between these layers. With older devices, the loss of anatomy is harder to appreciate, making their annotation a more challenging task. In addition, it seems that distinguishing between stage 1 and other stages of RPD presented a challenge for both humans and the model, as reflected by the better agreement (moderate vs. poor) when RPD lesions of all stages are considered, and as reflected in the qualitative examples (Figures 4 and 5). Despite the difficulty this dataset presents, the model achieved performance that is either beyond human performance (drusen, stage 1 RPD) or close to human performance (RPD stages 2,3, and 4; all RPD stages combined).

Of note, it is possible that the test set, chosen randomly, was challenging to annotate. For comparison, the intergrader agreement between the two graders who segmented the training and validation set was also calculated and was higher than that achieved for the test set. It was 0.94 (95% CI 0.91, 0.95) for drusen area, 0.67 (95% CI 0.6, 0.73) for RPD area when all stages were considered, 0.34 (95% CI 0.2, 0.45) for stage 1 RPD, and 0.72 (95% CI 0.67, 0.76) for stages 2, 3, and 4 RPD. If the test set was more challenging for humans, it can be implied that it was more challenging for a DL model, and better performance is expected on less challenging datasets. However, the difference in ICC scores may have resulted from higher consistency among the two graders who annotated the training and validation set separately from the three who graded the test set.

Similar findings were seen in the FROC curves. The model achieved sensitivity that is close to human performance, and in fact is similar to a senior retina expert, as evidenced by the overlapping confidence intervals seen in the plot. Given the complexity of grading these lesions, it is possible that more graders, especially with less experience, would have fared worse than the model. Of note is the improved sensitivity of the model with the increased number of average false positive lesions per B-scan, the importance of which depends on the settings under which the model is used. For example, we intend to use it to identify participants in the UKBB with RPD for further research, including genetic analysis, where quantification is key. To that end, a small number of false positives (i.e., high specificity) is required. Other settings might emphasize sensitivity over specificity, for example when the identification of patients with RPD is required for screening purposes. In such cases, a higher number of false positives might be allowed (especially if human validation is involved), enabling higher sensitivities for the model (almost 100% for drusen and 80% for RPD).

Another point to consider is the fact that the segmentation model was tested on a B-scan level. The FROC curves present an average number of false positives per scan. As with any average, some B-scans will fare better than others. When the model is deployed to whole volumes (eyes) to quantify RPD and drusen on a volumetric level, such inaccuracies may be less prominent.

The Dice scores (presented in Supplementary Table 1) reflect slightly better intergrader performance than model performance for all features (ranging from 0.06 for stage 1 RPD to 0.16 for all-stage RPD). Of note, both human and model performance as reflected in the Dice score were poor. As was shown in a recent publication by an international consortium of medical image analysis experts, Dice score is not an appropriate metric for small structures in images, since a single-pixel difference between two predictions can have a large impact on the metric difference.[47] Given the small size of the lesions graded in our work we utilized FROC as the primary metric to assess the performance of the segmentation model.

Thus far, only two studies have described the use of machine learning solutions for the automatic detection of RPD on OCT. In the first, by Saha et al.,[25] the authors trained several deep learning models to detect RPD, intraretinal hyperreflective foci, and hyporeflective foci within drusen. Although the model's performance was good for the detection of RPD (sensitivity of 79-96%, specificity 65-92%, AUC 0.91-0.94, accuracy 80-86%) all models were classifiers. Therefore, the output was binary, and quantification of the lesion area was not possible.

In the second study, by Mishra et al.,[26] the authors chose a different approach, whereby retinal layers associated with drusen and RPD were automatically segmented in SD-OCT images along with other retinal layers. The methodology involved a combination of a graph-based approach based on the Deep Learning - Shortest Path (DL-SP) algorithm on 2D OCT B-scan images. In that regard, drusen and RPD were considered types of layers - the former where the RPE layer is undulating, and the latter with undulation of the EZ. This technique presents several problems. First, undulations of these layers are not specific to drusen and RPD and may result from other pathologies. The extent to which this model can handle other pathologies and differentiate between them and the aforementioned lesions is unclear since the model was trained on 16 eyes with AMD only, and it is not clear if pathologies other than drusen (such as choroidal neovascularization) were included. In addition, while quantification of the lesions may be possible by calculating the inter-layer area for each of the lesions, this was not done in the study and it is unclear how accurate such a methodology would be.

Our group previously published a deep learning model for the segmentation of 13 features associated with neovascular and atrophic AMD. It achieved a lower ICC for drusen compared to the current study (0.381 ± 0.055, compared with 0.74 (95% CI 0.65, 0.82)) and lower sensitivity of under 40% for one average false-positive RPD (in comparison with 52% for the current model).[17] That is despite the former model being trained on a newer, higher resolution device (Topcon 3D OCT-2000 (Topcon, Tokyo, Japan)). That can be explained by the higher number of B-scans for both drusen and RPD used in the development of the current model, different DL architectures, and/or the quality of ground-truth annotations.

Previously published models are either not quantitative, have limited accuracy or cannot perform on the more challenging features of early generation SD-OCT devices. All three issues need to be addressed to accelerate our understanding of how RPD influence the pathogenesis of non-neovascular AMD, for which no licensed treatment exists. We are currently using transfer learning techniques to develop models for newer generation devices. Development of such models is less challenging given the higher resolution and easier delineation of small lesions, for both humans and algorithms.

In addition, this framework includes the only model that can accurately differentiate between drusen and RPD. Detection of both lesion types is needed to achieve an understanding of risk factors for RPD and drusen load by separating patients with RPD, RPD and drusen, drusen, and normal controls in future studies.

Our framework can be used in the future in treatment trials. For example, in the Laser Intervention in Early Stages of Age-Related Macular Degeneration (LEAD) study, aimed to evaluate the safety and efficacy of subthreshold nanosecond laser in intermediate AMD, it was found that such treatment may be inappropriate in patients with RPD compared to those without.[15] This suggests that treatments for those with RPD might need to be different from those with drusen, and as such it will be important to be able to accurately and quickly identify patients with or without RPD.

Our study has several limitations. a. It was trained on data from the UKBB. In the UKBB cohort, 94.6% of participants were of white ethnicity. This is similar to the national population of the same age range in the 2001 UK Census (94.5%) but slightly higher than in the 2011 Census (91.3%).[48] While these figures point at generalizability of the model to the UK population, it might not be generalizable to other populations with different ethnic and sociodemographic compositions; b. The performance of the different models was not evaluated in an external dataset, and might not generalize to datasets other than UKBB. c. We used different reference standards for the training and the test set. d. We evaluated each model within the suggested framework as separate steps and not as part of a continuous pipeline. e. We did not employ end-to-end training for this project, meaning that each model was optimized independently. f. The framework steps were trained on training data excluding questionable lesions. The ill-definition of this category, as well as the associated large intra- and inter-grader variability, prevent the definition of a reliable reference standard for training. This, however, might result in unpredictable behavior during inference if questionable lesions are present. Contingency strategies, such as uncertainty estimation or runtime failure detection, might help diminish a performance drop. f. We selected an experienced grader to serve as ground truth. However, as the results show, ground truth is difficult to determine as human agreement on this problem is low.

In conclusion, we present the first DL framework encompassing image quality assessment, differentiation of drusen and RPD from controls, and individual segmentation of these two phenotypes, with near-human performance. The application of this model in research settings and possibly in clinical settings will help further our understanding of RPD as a separate entity from drusen.

Acknowledgements

This study was sponsored by the EURETINA Retinal Medicine Clinical Research Grant.

**Figures**

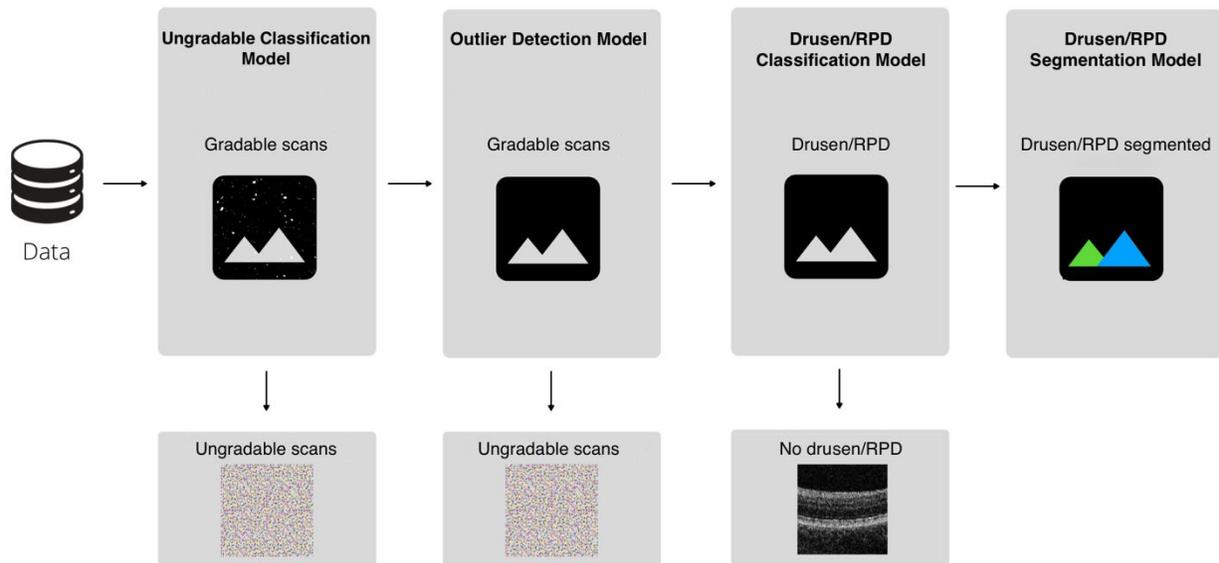

**Figure 1. Deep learning framework for the detection and quantification of conventional drusen and reticular pseudodrusen**

A classification algorithm classifies OCT volumes (on a volumetric (i.e., eye) level) into gradable or ungradable, and ungradable volumes are removed (Ungradable Classification Model). A deep ensemble model for out-of-distribution detection identifies volumes with out-of-distribution scans, which are then removed (Outlier Detection Model). Another classification model, the Drusen/RPD classifier, classifies the remaining volumes into those with either drusen or RPD versus controls. Controls are removed (Drusen/RPD Classification Model). Finally, an image segmentation algorithm segments RPD and drusen separately on a B-scan level (Drusen/RPD Segmentation Model). (RPD = reticular pseudodrusen).

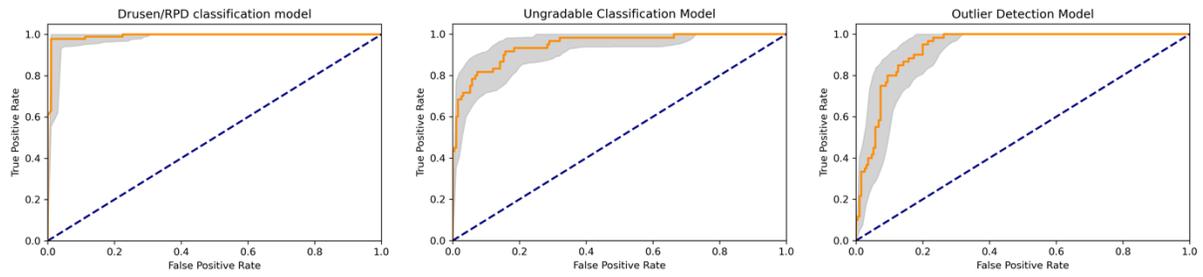

**Figure 2. Receiver operating characteristic (ROC) curves for the three classification models.** From left to right, the curves apply to the RPD/drusen vs. controls model, the Ungradable Classification Model, and the Outlier Detection Model. The orange line represents the models' sensitivity at different thresholds with the shaded area representing the 95% confidence interval.

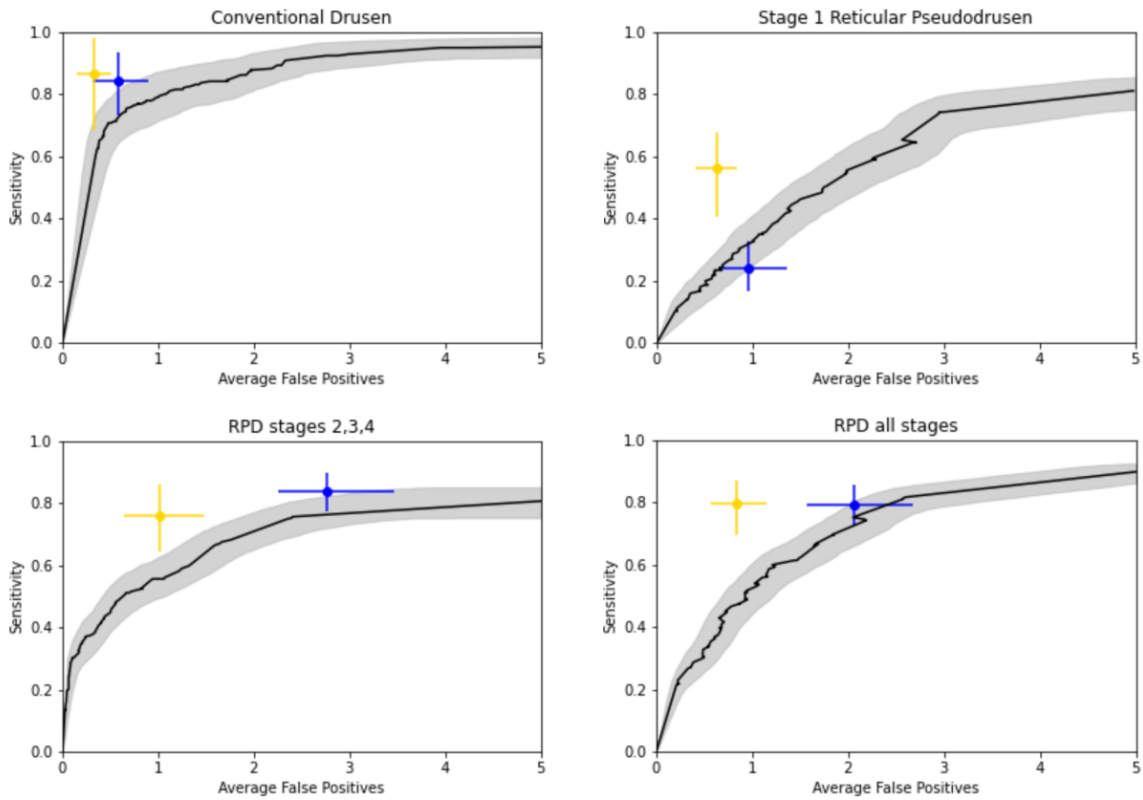

**Figure 3. Free response receiver operating characteristic (FROC) curves for drusen, stage 1 reticular pseudodrusen (RPD), stages 2,3, and 4 RPD, and all RPD stages combined, comparing the model to the ground truth graders.** The line represents model sensitivity at different thresholds, with the shaded area representing the 95% confidence interval, obtained by bootstrapping. The dots represent the two other graders, one represented in blue and the other in yellow, with error bars representing 95% confidence intervals.

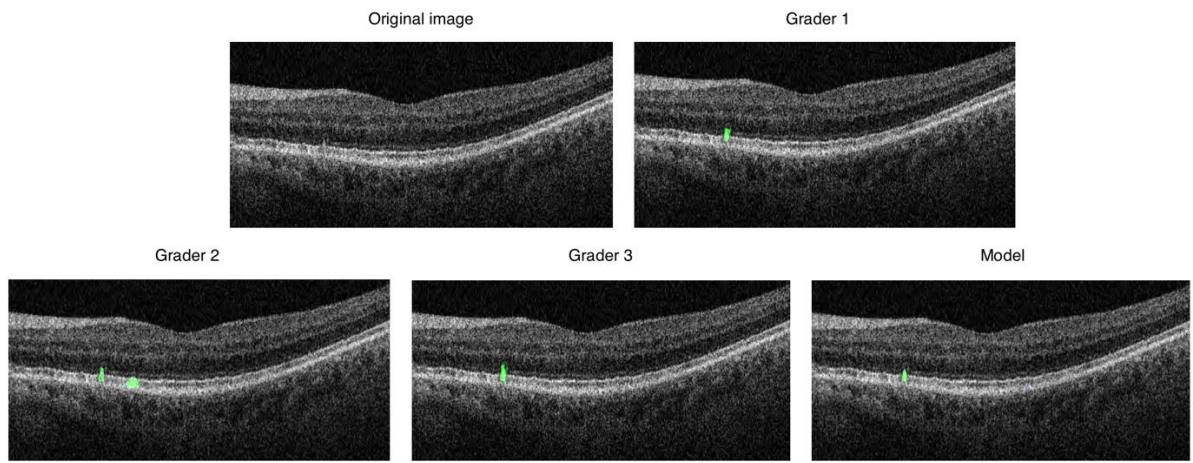

**Figure 4. Comparison of model and grader output, reticular pseudodrusen (RPD) stage 2, 3, or 4.** The green color represents stage 2, 3, or 4 RPD. Grader 1 was used as a reference standard against model performance and against other graders, presented in the FROC curves.

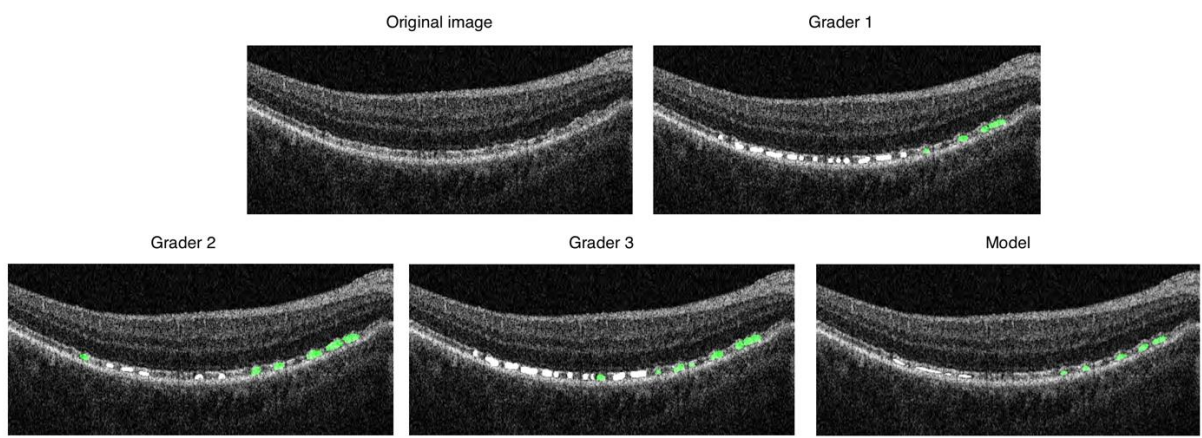

**Figure 5. Comparison of model and grader output, reticular pseudodrusen (RPD) stages 1 and RPD stages 2, 3, or 4.** Stage 1 RPD is represented in white and stage 2, 3, and 4 are represented in green. Grader 1 was used as a reference standard against model performance and against other graders, presented in the FROC curves.

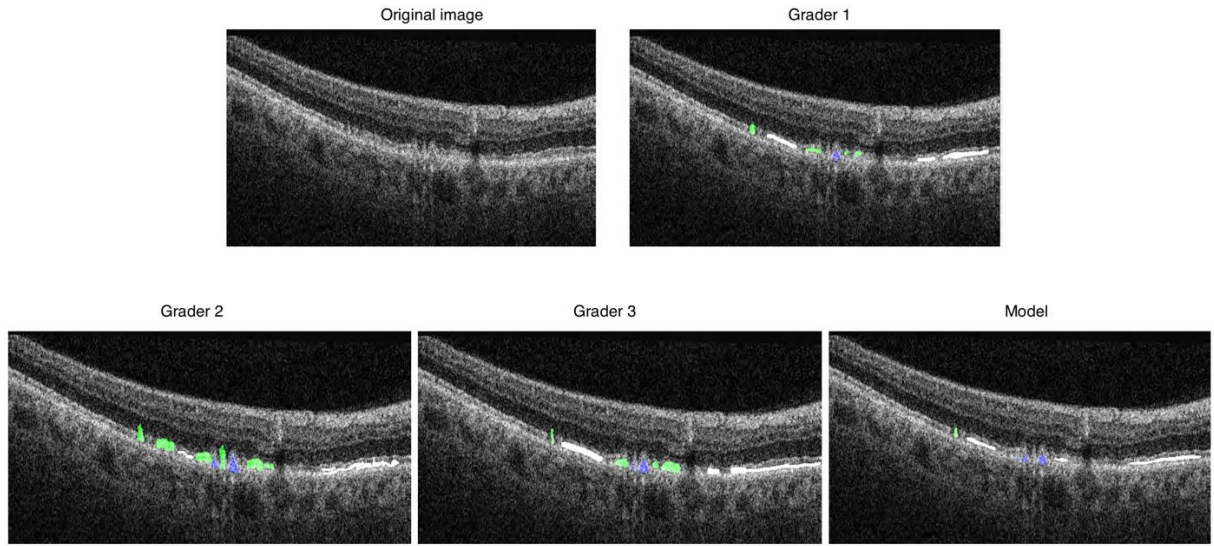

**Figure 6. Comparison of model and grader output, conventional drusen, reticular pseudodrusen (RPD) stages 1 and RPD stages 2, 3, or 4.** Drusen are represented in blue, stage 1 RPD in white, and stage 2, 3, and 4 are represented in green. Grader 1 was used as a reference standard against model performance and against other graders, presented in the FROC curves.

**Tables**

**Table 1 - Metrics obtained by the different classification models on the test set**

| Model | Sensitivity (%) | Specificity (%) | AUC | Kappa | Accuracy (%) | Area under the precision-recall curve |
|---|---|---|---|---|---|---|
| **Ungradable Classification Model** | 78.3 | 93.7 | 0.95 | 0.72 | 90.0 | 0.88 |
| **Outlier Detection Model** | 96.7 | 76.8 | 0.93 | 0.59 | 81.6 | 0.76 |
| **Drusen/RPD classification Model** | 97.8 | 99.0 | 0.99 | 0.97 | 98.4 | 0.99 |

**Table 2 - ICC Scores on the Test Set for the Model and Graders**

| Parameter | ICC model vs. grader 1 (95% CI) | ICC model vs. grader 2 (95% CI) | ICC model vs. grader 3 (95% CI) | **Mean ICC model vs. graders (95% CI)** | **ICC intergrader (95% CI)** |
|---|---|---|---|---|---|
| **Drusen area** | 0.68 (0.54, 0.77) | 0.71 (0.62, 0.77) | 0.82 (0.76, 0.86) | 0.74 (0.65, 0.82) | 0.69 (0.55, 0.78) |
| **RPD area - all stages** | 0.62 (0.5, 0.71) | 0.51 (0.24, 0.68) | 0.69 (0.52, 0.8) | 0.61 (0.50, 0.71) | 0.68 (0.61, 0.74) |
| **RPD stage 1 area** | 0.51 (0.4, 0.61) | 0.25 (0.12, 0.38) | 0.49 (0.38, 0.59) | 0.42 (0.25, 0.58) | 0.27 (0.18, 0.37) |
| **RPD stages 2,3,4 area** | 0.55 (0.37, 0.68) | 0.22 (-0.06, 0.46) | 0.43 (0.17, 0.61) | 0.4 (0.21, 0.59) | 0.48 (0.29, 0.62) |

**Supplementary material**

**Supplementary Table 1 – Mean Dice Scores on the Test Set between the model and each grader and between grader pairs**

|  | Model vs. grader 1 | Model vs. grader 2 | Model vs. grader 3 | **Average model vs. graders** | Grader 1 vs. grader 2 | Grader 1 vs. grader 3 | Grader 2 vs. grader 3 | **Average intergrader** |
|---|---|---|---|---|---|---|---|---|
| **Drusen area** | 0.29 | 0.33 | 0.33 | **0.32** | 0.37 | 0.51 | 0.41 | **0.43** |
| **RPD area - all stages** | 0.27 | 0.25 | 0.27 | **0.26** | 0.41 | 0.50 | 0.36 | **0.42** |
| **RPD stage 1 area** | 0.16 | 0.09 | 0.13 | **0.13** | 0.11 | 0.37 | 0.09 | **0.19** |
| **RPD stages 2,3,4 area** | 0.26 | 0.19 | 0.27 | **0.24** | 0.31 | 0.44 | 0.34 | **0.36** |

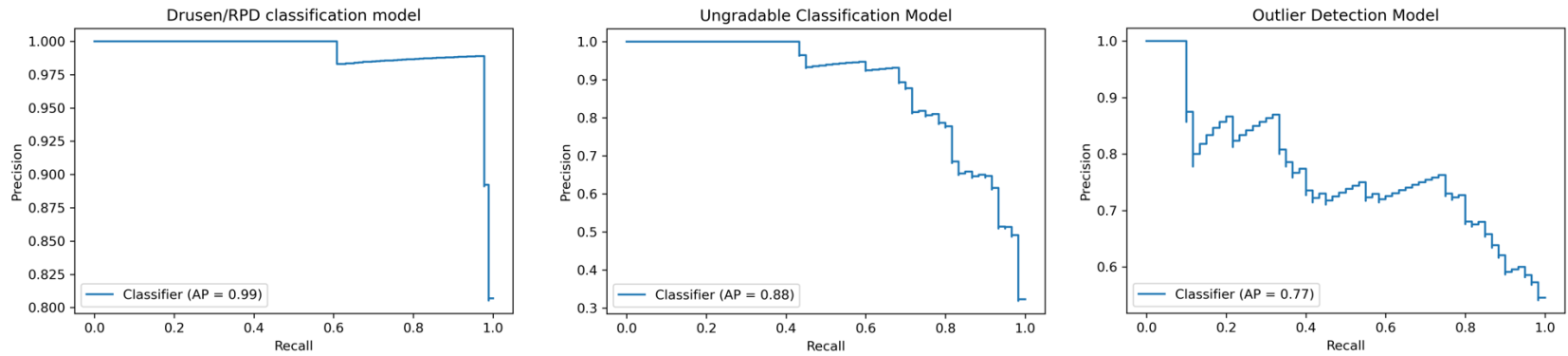

**Supplementary Figure 1**. **Precision-Recall curves for the different classification models.**

AP = Average precision